\documentclass[cup6b,epsf]{cupbook}
\usepackage{epsf}
\usepackage{epsfig}
\def\note #1]{{\bf #1]}}
\def\dd{{\rm d}}

\def\apj{{\it ApJ}}

\def\mnras{{\it MNRAS}}

\def\prl{{\it PRL}}
\def\jfm{{\it JFM}}

\begin{document}
\author{D. LYNDEN-BELL\\ Institute of Astronomy, The Observatories, \\
Madingley Road, Cambridge, CB3 0HA, U.K. \\
and Clare College, Cambridge, U.K.} 
\chapter{A Magic Electromagnetic Field}
{\it An electromagnetic field of simple algebraic structure is simply
derived.  It turns out to be the $G=0$ limit of the charged rotating
Kerr-Newman metrics.  These all have gyromagnetic ratio 2, the same as
the Dirac electron.  The charge and current distributions giving this
high gyromagnetic ratio have charges of both signs rotating at close
to the velocity of light.  

It is conjectured that something similar may occur in the quantum
electrodynamic charge distribution surrounding the point electron.}
\section{The Electromagnetic Field}
Away from charges and currents, both the electrostatic potential,
$\Phi$, and the magnetostatic potential, $\chi$, are harmonic. Thus $\Psi
= \Phi + i \chi$ satisfies 
$$
\nabla^2\Psi = 0\ .
$$
The solution obeying this equation everywhere -- except the origin -- and
tending to zero at infinity is $\Psi = q/r$, but if we move the origin
to ${\bf b}$ this solution becomes 
$$
\Psi = q \left/\sqrt{({\bf r} - {\bf b})^2 }\ \right.\ .
$$ 
This solution is harmonic whether $q$ and ${\bf b}$ are real or complex.

To ensure no magnetic monopole, term $q$ must be real, but we now
consider the possibility that ${\bf b} = i{\bf a}$ where ${\bf a}$ is
real so that ${\bf b}$ is pure imaginary.  Then we shall
have both an electric and a magnetic field with ${\bf F}={\bf E}+i{\bf
B} = - \mbox{\boldmath$\nabla$}\Psi$.  Without loss of generality we
may orient the $z$ axis along ${\bf a}$ so that 
\begin{equation}
\Psi = q\left/\left(R^2 + (z-ia)^2\right)^{1/2}\right. \hspace{1cm} {\rm
where}\hspace{1cm} R^2 = x^2+y^2\; . \label{eq:1}
\end{equation}
This expression will be harmonic except at singularities and branch
points.  The singularities lie at $R=a$ and $z=0$.  If we ask for no
branch points at infinity then we may take the cut defined by the disk
$z=0,\ R \leq a$, (but notice that we could take the cut around the sphere
$r=a,\ z \geq 0$ say).

We may evaluate $-\nabla \Psi $ to obtain
\begin{equation}
{\bf F} = {\bf E} + i {\bf B} = q({\bf r}-i{\bf a})\left/\left[({\bf r} -
i{\bf a})^2\right]^{3/2}\right.\ .\label{eq:2} 
\end{equation}
The total charge is clearly $q$ but the field also has a magnetic
dipole moment.  Indeed for $r>a$ we may use the Legendre polynomial
expansion of $\Psi$
\begin{equation}
\Psi = {q\over r}\sum^\infty_0 \left ({ia \over r}\right )^n P_n (\cos
\theta )\; . \label{eq:3} 
\end{equation}
Evidently all the $P_{2n}$ have real coefficients and all the
$P_{2n+1}$ have imaginary coefficients so the magnetic potential is
antisymmetrical about $z=0$ while the electric potential is
symmetrical.  Evidently the magnetic moment is the coefficient
of $iP_1$ which is $qa$ while the electric quadrupole moment is $qa^2$,
etc.  The relativistic invariants of the field are contained in 
$$F^2=E^2-B^2+2i{\bf E} \cdot {\bf B}
= q^2\left/\left[({\bf r}-i{\bf a})^2\right
]^2\right. \ .$$ 
Now $ \left [({\bf r}-i{\bf a})^2\right ]^2 $ is only imaginary if  
$({\bf r}-i{\bf a})^2 = \pm {1\over {\sqrt 2}\ } (1\pm i)\vert {\bf r} -
i{\bf a}\vert^2$ which occurs when
$\left (r^2-a^2\right )/(2{\bf r} \cdot {\bf a}) = \pm 1$
as then the real and imaginary parts are equal in
magnitude.  This condition may be rewritten $({\bf r}\pm {\bf a})^2 =
2a^2$  so  $E^2=B^2$ only on two spheres of radius $\sqrt{2}\ a$ centred
on   $({\bf r}= \pm {\bf a})$.  The circle in which they meet is the
ring $z=0,\ r=a$.  

Figure 1.1 illustrates where $\vert {\bf B}\vert >
\vert {\bf E}\vert $, etc.  ${\bf E}$ and ${\bf B}$ are perpendicular
when $({\bf r}-i{\bf a})^2 = r^2-a^2-2i{\bf a} \cdot {\bf r}$ is either
purely real or purely imaginary; i.e., on the sphere $r=a$, and the
plane $z=0$.  The Poynting vector is given by 
$${\bf F}^\ast \times {\bf F}= \left ({\bf E}-i{\bf B}\right ) \times
\left ({\bf E}+i{\bf B}\right ) = 2i {\bf E} \times {\bf B} = 2iq^2
{\bf a} \times {\bf r}\left/\left[\left({\bf r}-i{\bf a}\right )^2\right
]^3 \right.\; , $$ 
and the field energy density by $(8\pi)^{-1}{\bf F}^\ast \cdot {\bf F} =
(8\pi)^{-1} \left (E^2+B^2\right )$.  The velocity of the Lorentz frame in
which ${\bf E}$ and ${\bf B}$ are parallel is given by ${\bf v} =
c{\bf V}$ where 

\begin{eqnarray*}
{\bf V}\left/\left(1+V^2\right)\right. & = & {\bf E}\times{\bf
B}\left/\left(E^2+B^2\right)\right.
={\bf a}\times{\bf r}\left/\left(a^2+r^2\right)\right. \\
& = & {\bf F}^\ast\times{\bf F}
\left/\left(2i{\bf F} \cdot {\bf F}^\ast\right)\right. \; ; 
\end{eqnarray*}
squaring and solving for $V$ we find
\begin{eqnarray*}
V & = & \left [a^2+r^2 - \sqrt{\left (a^2+r^2\right )^2-4a^2R}\ \right]
\left/\left(2aR\right)\right. \\
& \\
& = &  2aR\left/\left[a^2+r^2+\sqrt{\left (a^2+r^2\right
)^2-4a^2R^2}\ \right ]\right.
 = aR\left/\left(a^2+\lambda\right )\right.\ ,\\
\end{eqnarray*}
where $$\lambda = {\scriptstyle{1\over 2}}\left [r^2-a^2 + \sqrt{\left
(r^2 -a^2\right )^2+4({\bf a} \cdot {\bf r})^2}\ \right ]\ , $$ 
is defined with the positive
root and $\mu$ is the same but for the negative root.  $\lambda$ and
$\mu$ are spheroidal coordinates.  Evidently $\Omega = V/R$ is
constant on the confocal spheroids, $\lambda =$ constant which have a
focal ring at the singularity.  (This result is due to
J. Gair.)

\begin{figure}[here]
\begin{center}
\leavevmode\epsfxsize=10cm \epsfbox{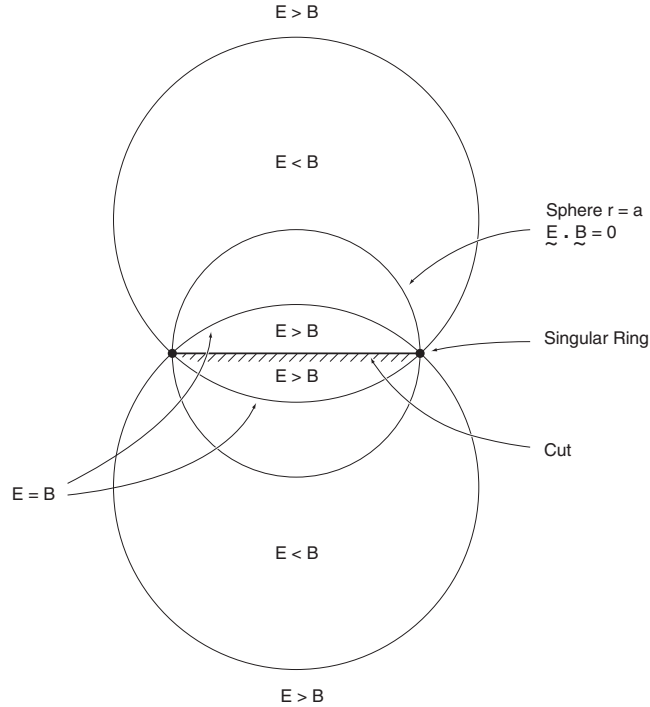}
\end{center}
\caption{Planar cut through the origin, orthogonal to the $z=0$ plane,
showing the delineation of regions of $E>B$ and $E<B$, for the 
potential given by eq. (1.1).}
\label{fig:one}
\end{figure}

On the cut itself we have $R<a$ and $z=0+$.  
$${\bf E}+i{\bf B} = q\left ({\bf R}-i{\bf a}\right )\left/i\left(a^2-R^2\right
)^{3/2}\right. = -q\left ({\bf a} + i{\bf R}\right
)\left/\left(a^2-R^2 \right )^{3/2}   \right.\; . $$   
This gives an electric field vertically down into the disc and a
magnetic field parallel to the disk surface for $R<a$ as though the
disk has  a Meisner effect.  The corresponding charge density on the
symmetry plane is 
$$\sigma = - \left (q\left/2\pi\right.\right ) a\left (a^2
-R^2\right )^{-3/2}\; . $$
This charge density gives a divergent total charge but that divergence
is cancelled by a ring of opposite charge on the edge which leaves the
total charge not `negative' but `positive' $+q$.  The total charge at
axial distance less than $R$ is $Q(<R)=-q\left
[a(a^2-R^2)^{-1/2}-1\right ],\ R<a$.  From the discontinuity in the
${\bf B}$ field across the cut we find $4\pi J_\phi =
-2qR(a^2-R^2)^{-3/2}$.  This corresponds to the charge density given
above rotating with angular velocity $\Omega = c/a$, reaching
the velocity of light at the singularity.  Again its effect is
reversed by a ring current at the edge.  The fields are illustrated in
Figures 1.2 and 1.3. 
\begin{figure}
\begin{center}
        \epsfig{file=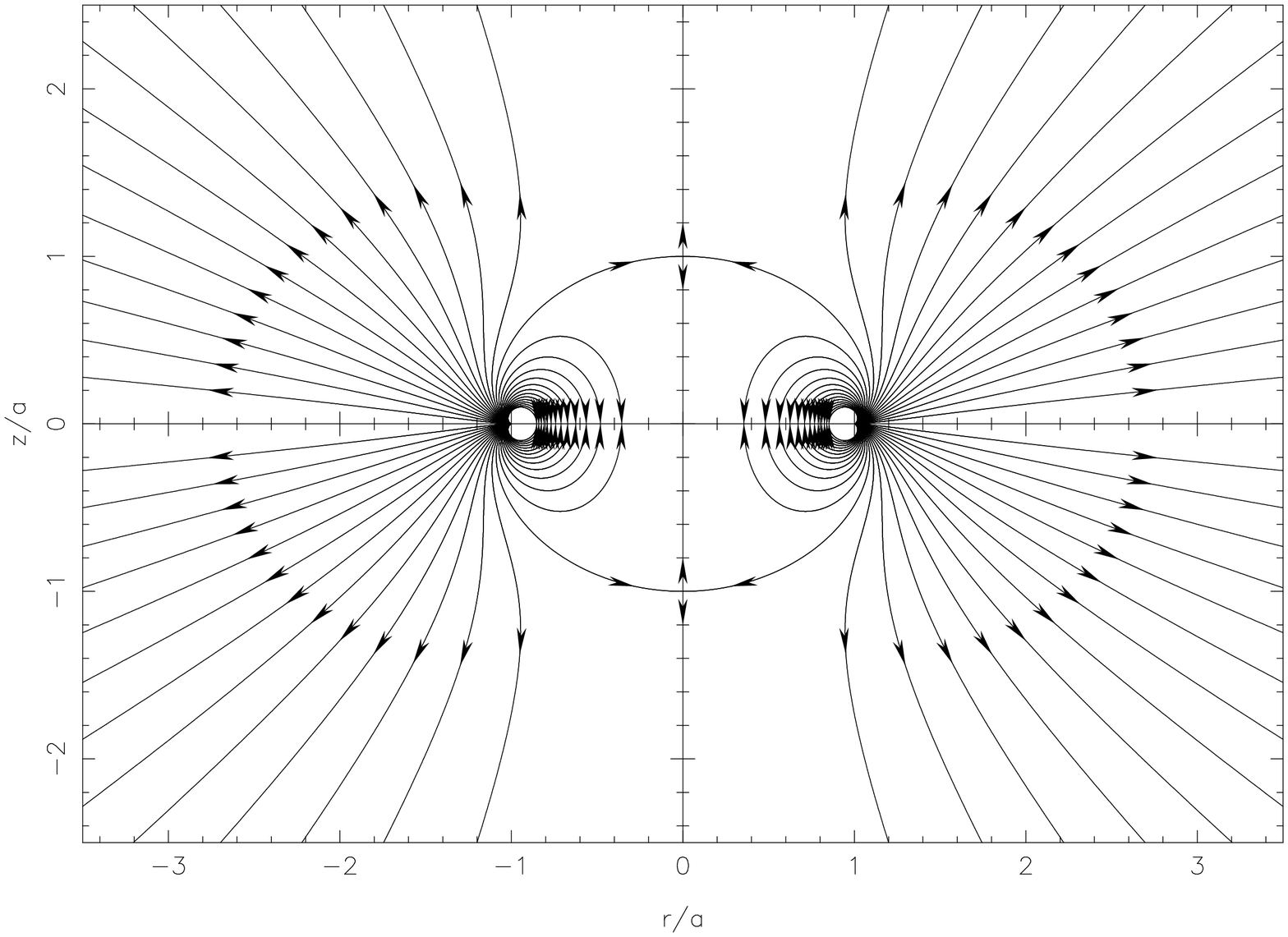, width=110mm}
        \caption{A plot of electric field lines for the potential 
given by eq. (1.1).\label{}}
        \vspace{10mm}
        \epsfig{file=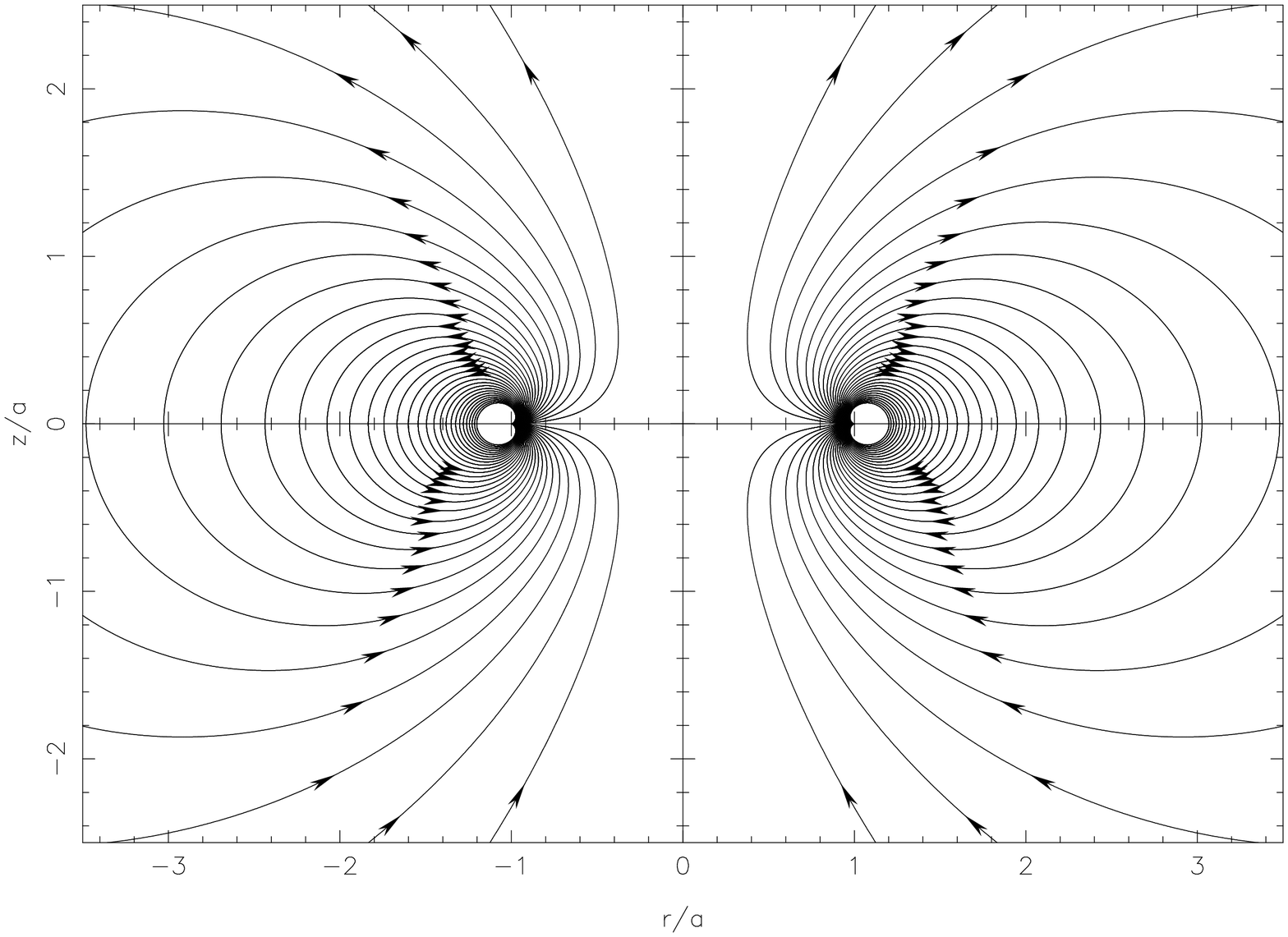, width=110mm}
        \caption{A plot of magnetic field lines for the potential 
given by eq. (1.1).\label{}}
\end{center}
\end{figure}

\section{The connection to Kerr's metric and the electron}
A much more complicated but more intriguing derivation of the above
results is to take the Kerr (1963) metric of a black hole of mass $m$
and \linebreak angular momentum $mac$.  Then complexify it following Newman
(1973) to get the Kerr-Newman metric of charge $q$, [Newman
et.al. (1965)].  Finally, take the limit with $G\rightarrow 0$ leaving
the charge and the moment corresponding to `$a$' but now in flat
space.  The resultant electromagnetic field is exactly that derived
and discussed above, [Pekeris \& Frankowski (1987)].  Carter (1968a)
showed that all the Kerr-Newman 
metrics had the same gyromagnetic ratio as the Dirac electron.  Does
this mean that there is some relationship between the charge
distribution of the Kerr-Newman metric and the charge distribution of
the quantum electrodynamic field of a point electron?  

Classical models of the electron had a problem over the gyromagnetic
ratio.  Even if all the charge were confined to a ring rotating at
close to the velocity of light the magnetic moment generated gives a
gyromagnetic ratio of one rather than the electron's value of 2.0023193044.   
It is of some interest to gain an understanding as to how the
Kerr-Newman metric does it.  The answer is that the charge
distribution is not all of one sign.  In fact a circular current
dipole of two rings of opposite charge rotating uniformly
 about their
common axis gives a net magnetic moment but no net charge.  The way
our electromagnetic field gets its large magnetic dipole moment per
unit net charge is that its much larger internal charges are of
opposite signs but rotate together giving a magnetic dipole with
relatively little net charge.  We show elsewhere that this is a
characteristic of relativistically rotating conductors!  
\section{Separability of Motion in the field}
Studies of separability of wave equations in the Kerr and Kerr-Newman
metrics [Carter (1968b), Teukolsky (1972, 1973), Chandrasekhar (1976), Page
(1976)] have shown that Dirac's equation is separable in these
metrics.  This of course implies that it is still separable in their
flat space limit as $G\rightarrow 0$.  The criterion for the
separability of Schr\"{o}dinger's equation in a real potential in
spheroidal coordinates is $\Phi = \left
[\zeta(\lambda)-\eta(\mu)\right]/(\lambda - \mu)$ [Morse and Feshback
(1953)].  Here $\lambda$ and $\mu$ are spheroidal coordinates and
$\zeta,\eta$ are arbitrary functions of their arguments.  

The field that we derived so simply above is rewritten in spheroidal
coordinates as follows:  $\lambda$ and $\mu$ are the roots for $\tau$
of the quadratic 
$${x^2+y^2 \over a^2+\tau} + {z^2 \over \tau} = 1\ ,$$
where $x^2+y^2 = R^2 = \left (\lambda + a^2\right )\ \left (\mu +
a^2\right )/a^2$ and the metric is 
$$\dd s^2=\dd x^2+\dd y^2+\dd z^2 
= {\lambda - \mu \over 4\lambda \left (\lambda +
a^2\right )}\dd \lambda^2 + {\lambda - \mu \over 4\mu \left (\mu + a^2
\right )} \dd \mu^2 + R^2\dd \phi^2\ .$$
To compare to Kerr's metric one uses the quasi-spherical form of
spheroidal coordinates $\widetilde{r}^2 = \sqrt{\lambda}\, ,\ \mu = -a^2 \cos^2 \vartheta,\ z=\widetilde{r}\cos \vartheta$.  Note however that $\widetilde{r}$
is constant on spheroids and $\widetilde{r} =  0$ is the disc $z=0,\ R\leq
a$.  Also $\vartheta$ is not the $\theta$ of spherical polar
coordinates but is constant in hyperboloids.  Thus
$$\dd s^2=\left(\widetilde{r}^2+a^2\cos^2\vartheta\right)\left/\right.\left(\widetilde{r}^2+a^2\right) \dd \widetilde{r}^2
+ \left(\widetilde{r}^2+a^2\cos^2\vartheta\right )\dd \vartheta^2 +$$
$$ + \left(\widetilde{r}^2+a^2\right )
\sin^2 \vartheta \dd \phi^2\ .$$   
In spheroidal coordinates our potential $\Psi = q/\sqrt{\left ({\bf
r}-i{\bf a}\right )^2}\ $  takes the simple forms
$$\Psi = q\left/\left(\sqrt{\lambda}\ -i\sqrt{-\mu}\ \right )\right. = q
{\sqrt{\lambda}\ + i \sqrt{-\mu}\ \over \lambda - \mu} = 
{q \over \widetilde{r} - ia \cos \vartheta}\ .$$
The second of these forms is exactly of the right type for
separability of the Schr\"{o}dinger equation but the similarity is 
partly misleading for Schr\"{o}dinger's equation only separates in an
electrostatic potential of that form.  When the imaginary (magnetic)
part is added Schr\"{o}dinger's equation no longer separates although
the Klein-Gordon equation now does separate (which it does not with
only the electrostatic part).  For a derivation and explanation of
these results see Lynden-Bell (2000).  

Systems with the same charge distribution but less magnetic field are
given by taking $\psi = \alpha \Psi +(1-\alpha)\Psi^\ast$ for $\alpha
<1$.  These magnetic fields are then multiplied by $2\alpha -1$.
These are weighted superpositions of discs rotating forwards and
backwards so the net rotation is less fast and $\alpha = 1/2 $ is
static.  These fields lose the magic of separability.  For the other
charge \& current distributions with that property see Lynden-Bell
(2000).
\section{Eulogy}
In closing, let me say that I still do not know the answer to the
problem discussed in my joint paper with Douglas [Gough \&
Lynden-Bell (1968)], i.e., ``How {\it do} turbulent fluids with angular
momentum like to rotate?''  \linebreak 
Nevertheless, I never expected to know the
internal rotation of the Sun within my lifetime and I have immense
admiration for Douglas -- and the helioseismic fraternity -- for having
persisted in analysing solar pulsations until that became possible.
Such is the real meat of good science.

\end{document}